\documentstyle[aps]{revtex}
\def\boldmu{\mbox{\boldmath $\mu$}}
\def\boldchi{\mbox{\boldmath $\chi$}}
\def\C{{\cal C}}
\def\H{{\cal H}}
\def\P{{\cal P}}
\def\U{{\cal U}}
\def\R{{\cal R}}
\def\S{{\cal S}}
\input epsf

\begin{document}
\draft

\title{THERMODYNAMIC LIMIT FOR DIPOLAR MEDIA\\}
\author{S. Banerjee\thanks{Electronic address: shubho@ipst.umd.edu}$^{1,2}$,
R. B. Griffiths$^1$, and M. Widom$^1$,}
\address{$^1$Department of Physics, Carnegie Mellon University,
Pittsburgh, Pa. 15213\\
$^2$Institute for Physical Science and Technology, University of Maryland,
College Park, Md. 20740~\thanks{Present address}}

\date{\today}
\maketitle

\begin{abstract}
	We prove existence of a shape and boundary condition
independent thermodynamic limit for fluids and solids of identical
particles with electric or magnetic dipole moments. Our result applies
to fluids of hard core particles, to dipolar soft spheres
and Stockmayer fluids, to disordered solid composites, and to regular 
crystal lattices. In
addition to their permanent dipole moments, particles may further 
polarize each other. Classical and quantum models are treated. Shape 
independence depends on the reduction in free energy
accomplished by domain formation, so our proof applies only in
the case of zero applied field. Existence of a thermodynamic limit
implies texture formation in spontaneously magnetized liquids and
disordered solids analogous to domain formation in crystalline solids.
\end{abstract}
\vspace{1cm}


\section{Introduction}
	Thermodynamics normally assumes a free energy density $F/V$
exists and is independent of system volume $V$ and shape.
Verification of these properties is impeded by the explicit dependence
of the partition function $Z=\exp{(-F/k_B T)}$ on these very
quantities.  Ruelle~\cite{dr} and Fisher~\cite{mef} proved the existence
of thermodynamic limits for a large class of fluids and solids with
interactions that fall off faster than $r^{-3}$ at large separation.
For such systems the free energy contains a boundary independent, {\em
extensive} (proportional to system volume) component and a boundary
dependent, {\em sub-extensive} (less than proportional to system
volume) remainder. Consequently, in the limit of infinite volume the
free energy density approaches a finite, boundary independent, limit.

The interaction energy between dipoles falls off
precisely as $r^{-3}$, seriously complicating the thermodynamic limit.
Volume integrals of this interaction (required to calculate
the total interaction energy $\H$) converge only conditionally because
the power of $r$ with which the interaction decays matches the
dimensionality of space. This paper considers systems with electric 
or magnetic dipole interactions. The electric and magnetic cases resemble 
each other closely. For convenience we carry out our discussion in the 
context of magnetism, then address electric analogues near the end of 
the paper.

	Long-ranged dipole interactions may create shape dependent
internal demagnetizing fields that increase the system free energy.
Boundary conditions on the surfaces may influence the strength of
these demagnetizing fields. The reduction in demagnetization energy
when uniformly magnetized regions break into smaller domains is the
key to the very existence of a thermodynamic limit in zero magnetic
field.  Griffiths~\cite{rbg} used the reversal of magnetization in a
domain to prove existence of a thermodynamic limit independent of
shape for dipolar lattices. We generalize that proof to include fluids
and disordered solids.  Certain conditions are required on the
``residual interaction'' $\H^R$, defined as the total interaction
energy $\H$ minus the magnetic interaction energy $\H^M$.  Our
proof, like Griffiths' original one, is valid only for zero applied
field because of its reliance on magnetization reversal in domains.

	The following section of our paper describes the origin of
demagnetizing fields, leading to a shape dependent free energy in the 
presence of an applied field. In section~\ref{shape_dependence}  we 
conjecture a simple functional form of the free energy, absorbing all 
shape dependence into a demagnetizing energy, implying a conventional 
thermodynamic limit for the remaining part of the free energy. 
Section~\ref{nofield} describes how the system achieves a  
thermodynamic limit in zero field.
Then, in section~\ref{main_proof}, we outline the formal thermodynamic
limit proof, which relies on upper and lower bounds on the free energy. 
We illustrate these bounds for stable and tempered systems in
section~\ref{stable_tempered}. Section~\ref{dipolar_systems} discusses
the difficulty dipolar interactions cause due to their lack of
tempering, and how that difficulty may be overcome.
Section~\ref{models} extends the proof to a variety of interesting
specific models, starting with identical hard core particles,
then treating dipolar soft spheres,
the Stockmayer fluid and polarizable particles. We treat both classical and
quantum versions of all these models. Section~\ref{electric} addresses
the analogous problems for electric dipoles. Finally, in
section~\ref{conclusion} we summarize our results, discuss some
observations about the implications of a thermodynamic limit for
spontaneously polarized liquids, and conclude with some interesting
dipolar systems which lack a thermodynamic limit.

\section{Demagnetizing fields}
\label{demagfield}

Long range dipole interactions create demagnetizing fields that cause
the shape and boundary condition {\em dependence} of free energy in
the presence of an applied field.  In zero applied field the
demagnetizing field must be handled with care to prove the shape and
boundary condition {\em independence} of free energy. This section
describes qualitatively why materials lack a shape independent
thermodynamic limit in an applied field, and how the limit is restored
in the absence of applied fields. We conjecture a modified form of
thermodynamic limit that may hold in an applied field, with all the
shape dependence restricted to an effective internal field plus an
explicit demagnetizing energy term.

Volume and surface distributions of magnetic poles cause 
demagnetizing fields. Consider a sample of magnetic material contained
in a region of space of volume $V$. Let ${\bf M}({\bf r})$ be the spatially
varying magnetization in the sample. Poles arise at surfaces wherever the
magnetization has a component normal to the surface $S$, as indicated in
figures~\ref{f1}a and~\ref{f1}b. A magnetization with non-zero divergence 
produces a charge density in the bulk. The demagnetizing field takes the form
\begin{equation}
\label{demag-field}
{\bf H}_D({\bf r})= \int_{S} d^2 {\bf r'}~ 
({\bf M}({\bf r'}) \cdot  {\bf \hat{n}}({\bf r'}))
{ {\bf r-r'} \over |{\bf r-r'}|^3}
-\int_{V} d^3 {\bf r'}~
({\nabla}' \cdot{\bf M}({\bf r'}))
{{\bf r-r'} \over |{\bf r-r'}|^3}
\end{equation}
where ${\bf {\hat n}}({\bf r'})$ is the outward normal at any point on the
surface. The spatial arrangement of surface charges depends on sample
shape, and the divergence of ${\bf M}({\bf r})$ in the interior depends on the
magnetization texture, so ${\bf H}_D({\bf r})$ is a function of sample shape
and magnetization texture. The demagnetizing energy~\cite{wfb}
\begin{equation}
\label{energy}
E_D =-{1\over{2}} \int d^3 {\bf r}~{\bf H}_D({\bf r}) \cdot {\bf M}({\bf r})=
{1\over{8 \pi}}\int d^3 {\bf r}~|{\bf H}_D({\bf r})|^2 \ge 0
\end{equation}
depends explicitly on the shape and magnetization texture of the
system through the demagnetizing field ${\bf H}_D({\bf r})$.

In the special case of magnetization uniform throughout the sample,
the demagnetizing field ${\bf H}_D({\bf r})$ comes only from the surface,
because the divergence term in equation~(\ref{demag-field}) vanishes.
However, ${\bf H}_D({\bf r})$ does not vanish as volume increases at fixed
shape. This is because the $1/r^2$ fall-off of the field from each
surface charge is exactly offset by the $r^2$ growth of surface area,
and hence the number of surface charges. As a result, ${\bf H}_D({\bf r})$ is
independent of the volume and the demagnetizing energy $E_D$ is extensive.

For the special case of a uniformly magnetized ellipsoid, ${\bf H}_D$
is constant within the ellipsoid and equals
\begin{equation}
\label{HD}
{\bf H}_D = - 4\pi {\bf D} \cdot {\bf M},
\end{equation}
where the tensor ${\bf D}$ is the demagnetizing factor of the
ellipsoid~\cite{wfb}. ${\bf D}$ is non-negative definite, and its
trace equals 1. When the magnetization lies along a principal axis of
the ellipsoid, ${\bf D}$ is simply replaced by one of its eigenvalues
$0 \le D \le 1$.  For a magnetization parallel to a highly elongated
needle shape, the demagnetizing factor $D=0$ because the
surface poles appear only on the tips which are small and far removed
from the bulk. Another special limit is that of magnetization normal to a
flat pancake shape. This yields the maximum demagnetizing effect,
since the surface poles appear on a large surface close to the bulk,
so $D=1$ in this case.

\subsection{Shape dependence in a field}
\label{shape_dependence}

When a system is placed in an external field ${\bf H}_0$, surface poles
arise because the internal magnetization tends to align with the applied 
field. There are two important contributions to the resulting shape 
dependence of the free energy. One is the explicit shape dependent 
energy~(\ref{energy}), the other is due to the shape dependence of the 
internal field
\begin{equation}
\label{H_i}
{\bf H}={\bf~H}_0+{\bf H}_D.  
\end{equation}
For highly elongated sample shapes, in the absence of demagnetizing
effects, we expect a thermodynamic limit for the free energy. Define
$f_{int}({\bf H}_0)$ as the free energy per unit volume of a system in the
limit as total volume $V\rightarrow \infty$.  The limit must be taken
within ellipsoidal shapes for which the length parallel to the field 
${\bf H}_0$ grows faster than the orthogonal directions. Because there 
are no demagnetizing effects present, we call this free energy density 
the {\em intrinsic} free energy density in a field.

For more general shapes ${\bf H}_D$ is non-zero, and may vary in
space. We conjecture that the shape dependent free energy
$F_{shape}$ may be expressed in terms of the intrinsic free energy
density as
\begin{equation}
\label{conjecture2}
F_{shape}({\bf H}_0) = \int d^3{\bf r} ~f_{int}({\bf H}({\bf r})) ~+~E_D
\end{equation}
up to corrections that grow less rapidly than the volume. 
Equation~(\ref{H_i}) gives the internal field {\bf H} and
equation (\ref{energy}) gives $E_D$.

Shape dependence of the free energy implies shape dependence of the
measured paramagnetic susceptibility. Assume that the magnetization 
${\bf M}$ is related to the
internal field ${\bf H}$ by an {\em intrinsic} (volume and shape
independent) {\em linear} susceptibility $\chi_{int}$ according to
\begin{equation}
\label{chi_i}
{\bf M} = \chi_{int} {\bf H}.
\end{equation}
Consider applying an external field ${\bf H}_0$ parallel to a
principal axis of an ellipsoidal sample.  Because of the demagnetizing
effect, the internal field {\bf H} is weaker than the applied field.
Eliminating ${\bf H}_D$ and ${\bf H}$ between equations (\ref{HD}),
(\ref{H_i}), and (\ref{chi_i}) yields the shape dependent {\em
measured} susceptibility
\begin{equation}
\label{chi_s}  
{\bf M} = \chi_{shape} ~{\bf H}_0
\end{equation}
where
\begin{equation}
\label{chi-shape}
 \chi_{shape} = {\chi_{int}\over{1+4 \pi D \chi_{int}}}.
\end{equation}
Note that the measured $\chi_{shape}$ has a maximum value equal to
$\chi_{int}$ when ${\bf H}_0$ is parallel to the long axis of a highly 
prolate needle-shaped ellipsoid. For any other geometry the demagnetizing
effect reduces the measured susceptibility.

\subsection{Shape independence in zero field}  
\label{nofield}

Now consider a ferromagnetic material in zero applied field. If the
magnetization were constant (Fig.~\ref{f1}a), surface poles
would create shape dependent demagnetizing fields and raise the energy as
described in equation~(\ref{energy}). A uniformly magnetized body lacks a 
shape independent thermodynamic limit!

Alternative magnetization configurations reduce the demagnetizing
energy. One possibility (Fig.~\ref{f1}b) reverses magnetization in
subregions so that the fields from surface poles tend to cancel.
Another possibility (Fig.~\ref{f1}c) rotates the magnetization so that
it is always tangent to the surface.  In each case, the reduction in
energy is proportional to the system volume $L^3$, where $L$ is a
typical linear dimension. The energy increase arising at a sharp
domain wall (Fig.~\ref{f1}b) should be proportional to the domain wall
area $L^2$.  The energy of a vortex line (Fig.~\ref{f1}c) should be
proportional to $L\log(L/a)$, with $a$ related to the vortex core
size. The magnetization texture may avoid a vortex by escaping into
the the third dimension~\cite{mermin} near the core. Such
textures contain either surface poles~\cite{ahroni} or point
defects~\cite{GD97}.  For sufficiently large systems, the extensive
$L^3$ reduction in the demagnetizing energy dominates the
sub-extensive defect energies, and domain wall, vortex, surface pole
or point defect formation is favored in that it lowers the free energy.

As we take the $V$ going to infinity limit, the demagnetizing energy 
density $E_D/V$ approaches zero for the most favorable magnetization,
i.e., the one which minimizes the energy. 
Such a magnetization permits a shape independent 
thermodynamic limit for a ferromagnet. The free energy density for an 
arbitrary shape with its nonuniform equilibrium magnetization texture 
equals the free energy density of a highly elongated
needle-shaped ellipsoid with uniform magnetization parallel to the
long axis, because $D$ tends to zero for the needle shape. When
calculating magnetic energies or free energies it may be convenient to
impose the needle-shape and assume uniform polarization. Alternatively,
``tin foil'' boundary conditions~\cite{leeuw} may be used to neutralize
the surface poles.

Why is zero applied magnetic field essential for a thermodynamic limit? In
an applied magnetic field the energy cost for flipping a domain
becomes proportional to the domain volume (and grows proportionally to
$L^3$) rather than the smaller domain wall, vortex or other defect energy.  
The most favorable magnetization texture now has a demagnetizing field, 
and the free energy re-acquires a shape dependence.

The above discussion explains how domain formation removes the
demagnetizing energy density $E_D/V$, permitting a shape independent 
{\it free energy} for zero field ferromagnets. This argument does not 
apply to the zero field {\em susceptibility} of a paramagnet. The shape 
dependent susceptibility $\chi_{shape}$ governs fluctuations in the average
magnetization of the entire sample.  When this average fluctuates from
zero, demagnetizing fields increase the free energy and oppose the
fluctuation. Reduced fluctuations imply a reduced susceptibility that
depends explicitly on shape through equation~(\ref{chi-shape}).

Still, we expect shape independent values of magnetic permeability
$\mu=1+4 \pi \chi_{int}$ (or dielectric constant $\epsilon$ in the 
case of electric polarizability). This can be understood by expressing
the permeability in terms of spatial integrals of correlation
functions~\cite{deutch,wertheim}. These correlation functions contain
short-ranged, shape independent, components, and long-ranged, shape
dependent components. The permeability depends only on the
short-ranged, shape independent, part of the correlation functions.

\section{Proof of the Thermodynamic Limit}
\label{main_proof}

This section explains how we prove thermodynamic limits. First, we
state required bounds on the free energy and explain how these bounds
are used to prove the existence of a thermodynamic limit. Then, we show
how to prove the necessary bounds on the free energy for classical systems 
which are stable and tempered. These sections are rather brief and formal,
and simply review methods introduced previously~\cite{dr,mef}. Then,
in section~\ref{dipolar_systems} we show how to treat systems which
include unstable and non-tempered dipole interactions.

\subsection{Conditions on the Free Energy}
\label{formal}

Consider an $N$ particle system contained in a region $\R$ of volume $V$. 
Taking the thermodynamic limit for the free energy means constructing a 
sequence of sufficiently regular regions~\cite{mef}, with increasing 
volume, so that the number of particles $N$ divided by $V$ approaches 
a definite value $\rho$ as the volume $V$ tends to infinity.
A limit is said to exist for the free energy density if the free energy $F$
divided by the system volume $V$ approaches a limiting value $f$ 
as the volume tends to infinity.
The requirement of regularity~\cite{mef} prevents
the regions $\R$ from getting too thin or constricted. We also introduce a
model-dependent density $\rho_c$ that ensures the particles can fit into
the available volume when $N/V$ is less than $\rho_c$ for sufficiently 
large finite $N$.

Two conditions on $F$ suffice to prove the thermodynamic
limit.

1) The free energy $F$ should 
satisfy the {\em lower bound}
\begin{equation}
\label{lower_bound}
F \ge V f_L(N/V)
\end{equation}
for $N/V < \rho_c$ where $f_L(\rho)$ is some finite valued function.

2) Consider a system composed of two subsystems, $1$ and $2$, containing
$N_1$ and $N_2$ particles, respectively. The particles in subsystem $1$ are 
confined in a region $\R_1$ with volume $V_1$ and those in 
subsystem $2$ are confined in a region $\R_2$ with volume
$V_2$. The two regions $\R_1$ and $\R_2$ are separated by a distance of
at least $d$ from each other.  Provided that $d\ge d_0$, for some 
fixed distance $d_0$, the free energy of the system should satisfy an {\em
upper bound}
\begin{equation}
\label{upper_bound}
F \le F_1 + F_2 + \Delta_{12},
\end{equation}
where $F_1$ and $F_2$ are the free energies of subsystems $1$ and $2$
in isolation and~\cite{temper}
\begin{equation}
\label{Delta}
\Delta_{12} \equiv (N_1 + N_2)^2 \omega_B/d^{3+\epsilon},
\end{equation}
with constants $\omega_B<\infty$ and $\epsilon>0$.

These bounds suffice for proving the existence and shape independence of the
thermodynamic limit. Break an arbitrarily shaped system into many smaller
subsystems. The upper bound~(\ref{upper_bound}) bounds the total free
energy in terms of the subsystem free energies. Because the upper bound applies
regardless of the relative positions of the subsystems, provided $d \ge d_0$,
the original system shape does not enter this bound on total free energy.
The lower bound~(\ref{lower_bound})
guarantees that the free energy density $F/V$ reaches a finite limit
as the total volume $V~\rightarrow~\infty$ with
$N/V~\rightarrow~\rho$. Because Fisher~\cite{mef} explains this method
in great generality, we need not reproduce his effort here.

\subsection{Classical Stable and Tempered Systems}
\label{stable_tempered}

The free energy of a classical system of $N$ identical particles in a
volume $V$ is $F = -k_B T \log{Z}$, where $Z$ is the partition function
\begin{equation}
\label{pf}
Z = {{1}\over{\Omega^N N!}} \int_V \prod_{i=1}^N 
d^3{\bf r}_i  d\Omega_i e^{- \H/k_B T}.
\end{equation}
The energy $\H$ is a function of particle center of mass positions
${\bf r}_i$, and internal coordinates $\Omega_i$. In the expression
for $Z$, $\Omega$ is the integral of $d \Omega_i$ over all its
possible values. Internal coordinates depend on the type of particle
and may include orientation of the particle and direction of
magnetization. For solids, particle center of mass positions and
particle orientations are fixed, and the principal remaining variable
is direction of magnetization.

We distinguish between two types of particle: superparamagnetic
particles~\cite{superparamagnetism}, for which the direction of
magnetization rotates independently of the particle axes; normal
particles, for which the direction of magnetization is fixed relative
to the particle axes. For normal particles, we do not include
direction of magnetization as an independent internal variable,
because it is a function of particle orientation. In practice,
superparamagnetic particles exhibit a ``blocking temperature'' below
which the direction of magnetization becomes locked to the particle
axes, and the particles become normal. In the specific models
discussed below, we assume we are below the blocking temperature
except where we explicitly invoke superparamagnetism.

Note the explicit dependence of $Z$ on the system shape through the
limits of integration for the ${\bf r}_i$ in equation (\ref{pf}). 
The free energy $F$
inherits this shape dependence. The conditions on $F$ stated in
section~\ref{formal} guarantee that the shape dependence is contained
entirely in a sub-extensive term. Achieving the desired lower and
upper bounds on free energy depends on properties of the interaction
energy $\H$. This section describes sufficient conditions to prove
each bound.

The lower bound~(\ref{lower_bound}) holds for potentials that are 
{\em stable} in the sense that
\begin{equation}
\label{stable}
\H \ge -N \omega_A
\end{equation}
with $\omega_A<\infty$ a constant. Just substitute the lower bound
(\ref{stable}) for $\H$ into the partition function (\ref{pf}) to
obtain the lower bound (\ref{lower_bound}) on $F$, with the function
\begin{equation}
\label{fL}
f_L(\rho)=\rho k_B T \log{\rho} - \rho \omega_A.
\end{equation}

To prove the upper bound~(\ref{upper_bound}), consider the interaction of
two subsystems separated by distance $d$ as described in
section~\ref{formal}. Write the total energy $\H$ in the form
\begin{equation}
\label{hamiltonian}
\H \equiv \H_1 + \H_2 + \H_{12},
\end{equation}
where $\H_1$ and $\H_2$ denote the energies of each system by itself,
and $\H_{12}$ is the interaction energy between the two subsystems.
The upper bound holds if the interaction $\H_{12}$ satisfies the {\em weak
tempering} condition~\cite{temper}
\begin{equation}
\label{tempered}
\H_{12} \le \Delta_{12}
\end{equation}
with $\Delta_{12}$ as defined in equation~(\ref{Delta}), for $d$ larger than
some constant $d_0$. Substitute
$\Delta_{12}$ for $\H_{12}$ in the total interaction
energy~(\ref{hamiltonian}) and evaluate the partition
function~(\ref{pf}) to derive the upper bound~(\ref{upper_bound}) on
$F$.

\subsection{Dipolar Systems}
\label{dipolar_systems}

The remainder of this paper considers systems whose Hamiltonians
include dipolar interactions in addition to stable, tempered
interactions of the type described above. The dipole interaction, by
itself, is neither stable nor tempered. In this section we explain how
additional repulsive interactions may stabilize the system, and how
the upper bound~(\ref{upper_bound}) may be proven despite the lack of
tempering. The ideas introduced here are applied to a wide variety of
specific models in section~\ref{models}.

Split the interaction energy ${\cal H}$ into two components:
\begin{equation}
\label{residual}
\H = \H^M+\H^R.
\end{equation}
The non-magnetic part of the interaction, $\H^R$, we call the residual
interaction. The magnetic interaction between the $N$ particles takes the
form~\cite{wfb}
\begin{equation}
\label{HM}
\H^M= \sum_{i<j=1}^N \int_{v_i} d^3 {\bf r}~ \int_{v_j} d^3 {\bf r}'
{\biggl \{}{{\bf M}({\bf r}) \cdot {\bf M}({\bf r}') \over |{\bf r-r'}|^3}-
{3({\bf M}({\bf r}) \cdot ({{\bf r-r'}}))~({\bf M}({\bf r'}) \cdot 
({ {\bf r-r'}})) \over |{\bf r-r'}|^5}{\biggl \}},  
\end{equation}
where $v_i$ and $v_j$ are the regions of space occupied by the
magnetic material of particle $i$ and $j$. The magnetization
distribution of the $i^{th}$ particle is ${\bf M}({\bf r})$ for ${\bf
r}$ inside $v_i$, and for the $j^{th}$ particle is ${\bf M}({\bf r'})$
for ${\bf r'}$ inside $v_j$. Implicitly, $\H^M$ depends on the
particle center of mass positions and the particle orientations
through $v_i$ and $v_j$, the regions of space occupied by the
particles. For superparamagnetic particles, the direction of
magnetization is an internal coordinate for each particle, while for
normal particles, the magnetization is determined by the particle
orientation. Thus, $\H^M$ is a function of particle positions 
$\{{\bf r}_i\}$ and internal coordinates $\{\Omega_i\}$.

For the moment we consider only permanent magnetization (polarizable
particles are discussed in section~\ref{polar}), and we assume the
magnetized volumes of the particles are non-overlapping.  The $1/r^3$
dependence of the magnetic interaction $\H^M$ violates tempering
because of its slow decay at long range, and risks violating stability
because of its divergence at short range.

We demand stability of the total interaction $\H$ to enforce the lower
bound~(\ref{lower_bound}) on the free energy. Because of the diverging
short-range magnetic attraction, we need residual interactions that
are sufficiently repulsive at short range to overcome the magnetic
attraction. Hard-core particles, and soft-core particles with energies
that diverge faster than $1/r^3$, satisfy this requirement, as we prove
later in section~\ref{models}.

To achieve the upper bound~(\ref{upper_bound}) on the free energy we 
demand that the residual interaction $\H^R$ be tempered
and we exploit symmetries (if present) to handle the non-tempered magnetic
interaction $\H^M$. Our strategy limits our proof to models possessing
the required symmetries and tempering of residual interactions. Models
lacking these characteristics may still possess a thermodynamic limit even
though we cannot prove it. The symmetries we require are broken by
applied magnetic fields.

Consider two subsystems such that the $N_1$ particles in region $\R_1$
are separated by at least a distance $d \ge d_0 > 0$ from the $N_2$
particles in region $\R_2$.  Let $\H_1$ and $\H_2$ be their respective
Hamiltonians. Define the interaction energy between the two subsystems
by
\begin{equation}
\H_{12} \equiv \H-\H_1-\H_2 = \H_{12}^M+\H_{12}^R.
\end{equation}	
Let $F(\lambda)$ be the free energy of the combined system when the
Hamiltonian is $\H_1+\H_2$ plus a scaled interaction $\lambda
\H_{12}$.  Because $F(\lambda)$ is a concave function (that is,
$F''(\lambda)\le 0$), it is bounded above by
\begin{equation}
F(\lambda) \le F(0)+\lambda F'(0),
\end{equation}
where the right side is a line tangent to the graph of $F(\lambda)$
at $\lambda=0$; here $F'(\lambda)$ and $F''(\lambda)$ are the first
and second derivatives. As a consequence, the free energy $F(1)$ of
the fully interacting system satisfies
the Gibbs inequality~\cite{falk}
\begin{equation}
\label{bogo}
F(1) \le F(0) + \langle \H_{12}^M+\H_{12}^R \rangle_{\lambda=0},
\end{equation}
where $F(0)=F_1+F_2$ is the free energy of the non-interacting subsystems,
and the classical ensemble average of any quantity ${\cal Q}$ takes the form
\begin{eqnarray}
\label{pert}
\langle {\cal Q} \rangle_{\lambda=0}
= {{1}\over{\Omega^{N_1+N_2} N_1! N_2! Z_1 Z_2 }}
\int_{V_1} \prod_{i=1}^{N_1} d^3{\bf r}_i d\Omega_i
\int_{V_2} \prod_{j=1}^{N_2} d^3{\bf r}_j d\Omega_j 
{\cal Q} e^{-(\H_1+\H_2)/k_B T}.
\end{eqnarray}
Because the residual interaction $\H^R$ is tempered, therefore $\H_{12}^R \le
\Delta_{12}$. Its ensemble average, likewise, is bounded above:
\begin{equation}
\label{mean_H12R}
\langle \H_{12}^R \rangle_{\lambda=0} \le \Delta_{12}.
\end{equation}
We will show that
\begin{equation}
\label{mean_H12M}
\langle~\H_{12}^M~\rangle_{\lambda=0} = 0.
\end{equation}
Combining the bound~(\ref{bogo}) with the ensemble
averages~(\ref{mean_H12R}) and~(\ref{mean_H12M}) proves the upper
bound~(\ref{upper_bound}) on the fully interacting free energy $F(1)$.

To establish~(\ref{mean_H12M}) we employ what we call a
$\theta$ operator, a map of the coordinates of a system onto themselves
in a one-to-one manner satisfying the following conditions. It leaves the 
center of mass position  ${\bf r}_i$ of each particle unchanged, it maps the 
internal coordinates $\Omega_i$ onto themselves in a way which leaves the 
integration measure $\prod_{i=1} d\Omega_i$ unchanged, and it  leaves the
Hamiltonian $\H$ invariant. In addition, when a system consists of two
subsystems and $\theta$ is applied to one but not the 
other, it reverses the sign of the magnetic interaction $\H_{12}^M$
between them. Specific models may or may not possess such an operator. 
When a system is stable and possesses a $\theta$ operator, we can obtain 
the desired upper and lower bounds on $F$ to prove a thermodynamic limit.

If a $\theta$ operator exists, it can be used  to establish~(\ref{mean_H12M})
in the following way. Set ${\cal Q}=\H_{12}^M$ in~(\ref{pert}), and consider 
the change of variables produced by applying the $\theta$ operator to 
subsystem $1$ but not subsystem $2$. This change preserves the integration
measure, but reverses the sign of the integrand, since $\H_1$ and $\H_2$ are
unaltered, but $\H_{12}^M$ changes sign. Consequently, the integral is equal
to its negative, so it is zero.

\section{Models}
\label{models}

	Section~\ref{main_proof} introduced a general strategy for
proving thermodynamic limits of permanently magnetized classical
particles. The following section applies that strategy  to a variety of
models. We start with identical hard core particles, then treat dipolar 
soft spheres such as Stockmayer fluids.  We then modify 
the proof to cover polarizable particles, and then treat quantum
systems. Depending on the particular system, the greater challenge may
lie in demonstrating the lower, or the upper, bound on $F$.

\subsection{Identical hard core particles}
\label{hard_core_particles}

Consider a collection of $N$ identical, uniformly magnetized, hard core 
normal particles of volume $v$, and fully  contained within a region 
of space $\R$ of volume $V$.  The magnetization ${\bf M}({\bf r})$  
is constant in magnitude for 
${\bf r}$ in volume $v_i$ of particle $i$  and vanishes when ${\bf r}$ 
is not inside a particle. Inside particle $i$ the direction of 
${\bf M}({\bf r})={\bf M}_i$ depends on the orientation $\Omega_i$ of the 
particle. We require that the region $\R$ have a regular shape~\cite{mef} 
and be large enough so that all particles  fit inside the region without 
overlapping. Thus, we restrict the number of particles so that the packing 
fraction $\phi \equiv {N v/ V}$ is not too large. In particular, we 
assume a packing fraction $\phi^*>0$ exists for which particles may be 
packed with any  $0 \le \phi \le \phi^*$ into any sufficiently large and 
regular volume.

Write the Hamiltonian as
\begin{equation}
\label{dhs}
\H= \H^M+\H^{HC}.
\end{equation}
The hard core interaction,
\begin{equation}
\label{HC}
\H^{HC}=\left\{ \begin{array}{ll}
0 & \mbox{if no particles overlap,} \\ \nonumber
+\infty & \mbox{otherwise},
\end{array}
        \right.
\end{equation}
prevents any overlap between particles. For non-overlapping configurations
the magnetic interaction $\H^M$ is as in~(\ref{HM}). For the special case
of hard core spheres the expression~(\ref{HM}) reduces to the simpler form
\begin{equation}
\label{HDD}
\H^M= \sum_{i<j=1}^N {\boldmu_i \cdot \boldmu_j-3(\boldmu_i  \cdot 
{\hat {\bf r}_{ij}})
(\boldmu_j  \cdot {\hat {\bf r}_{ij}}) \over {r_{ij}^3}}
\end{equation}
where $\boldmu_i$ and ${\bf r}_{i}$ are the dipole moment and position
of the $i^{th}$ particle; $\boldmu_i$ is the integral of ${\bf M}({\bf r})$
over the volume of the $i^{th}$ particle; ${\bf r}_{ij}=
{\bf r}_{i}-{\bf r}_{j}$, and ${\hat {\bf r}}_{ij}$ is the unit vector along 
${\bf r}_{ij}$.

The hard core interaction~(\ref{HC}) by itself provides an example of
a stable and tempered interaction. Since $\H^{HC} \ge 0$, it obeys the
stability condition~(\ref{stable}) with $\omega_A=0$.
Griffiths~\cite{rbg} proved the lower bound
\begin{equation}
\H^M \ge -{N \mu^2 \over{2 R^3}},
\end{equation}
for non-overlapping dipolar spheres with radius $R$, and dipole moment 
$\mu$, regardless of their positions and orientations. Because of the 
hard core repulsion~(\ref{HC}) we achieve stability~(\ref{stable}) with
$\omega_A=\mu^2/2R^3$ for dipolar hard spheres~(\ref{dhs}). The lower bound 
on $F$ follows as discussed in section~\ref{stable_tempered}. We now generalize
the proof of stability~(\ref{stable}), and thus a lower bound on $F$, to 
particles of all shapes.

To prove stability we make use of the positivity of field energy. Adding 
the magnetic self energy of each particle to $\H^M$ gives the total energy 
of the whole system, considered as one magnetization 
distribution,
\begin{equation}
\label{prior}
\H^T=\H^M+\sum_{i=1}^N E_i^{self}=-{1\over{2}} \int d^3 {\bf r}~
{\bf H}_D({\bf r}) \cdot {\bf M}({\bf r}).
\end{equation}
Here ${\bf H}_D({\bf r})$ is the field, due to all particles, defined
in equation~(\ref{demag-field}) and
\begin{equation}
\label{selfnrg}
E_i^{self} = - {{1}\over{2}} {\bf M}_i \cdot \int_{v_i} 
d^3 {\bf r}~{\bf H}_{iD}({\bf r}),
\end{equation}
where ${\bf H}_{iD}({\bf r})$ is the field from magnetization ${\bf M}_i$
of particle $i$ with volume $v_i$, obtained by substituting ${\bf M}_i$ for 
${\bf M}({\bf r})$ in equation~(\ref{demag-field}) and integrating  over the
surface and volume of particle $i$.

We use equation~(\ref{prior}) to place a lower bound on $\H^M$ by a
method similar to that of Griffiths~\cite{rbg}.  For any magnetization
distribution ${\bf M}({\bf r})$ and the field ${\bf H}_D({\bf r})$
caused by it
\begin{equation}
\label{identity}
-{1\over{2}} \int d^3 {\bf r}~{\bf H}_D({\bf r}) \cdot {\bf M}({\bf r})
={{1}\over{8 \pi}} \int d^3 {\bf r}~|{\bf H}_D({\bf r})|^2 \ge 0.
\end{equation}
Hence
\begin{equation}
\H^M+\sum_{i=1}^N E_i^{self} \ge 0.
\end{equation}
Brown~\cite{wfb} rewrites the self energy in~(\ref{selfnrg}) as
\begin{equation}
E_i^{self} = 2\pi \sum_{k,l} D_i^{kl} M_i^k M_i^l~~v_i,
\end{equation}
where ${\bf D}_i$ is the demagnetizing tensor of an ``equivalent
ellipsoid''; it exists for a particle of any shape, and $k$ and $l$ 
index the components of ${\bf D}_i$ and ${\bf M}$. Since ${\bf D}_i$ is
positive definite, with trace equal to 1,
\begin{equation}
E_i^{self} \le 2\pi M^2 v_i.
\end{equation}
Since all particles are identical, the magnetic interaction satisfies the 
lower bound
\begin{equation}
\label{stability}
\H^M \ge -2\pi M^2 N v.
\end{equation}
Thus we confirm 
stability~(\ref{stable}). The lower bound~(\ref{lower_bound}) on the
free energy follows with $\omega_A=2\pi M^2 v$. 

For proving an upper bound on the free energy, notice that the hard core
interaction~(\ref{HC}) is tempered, equation~(\ref{tempered}), with 
any $d_0>0$. We identify the hard core interaction~(\ref{HC}) as a
residual interaction~$\H^R$. The key to our proof of an upper bound in 
Section~\ref{dipolar_systems}
was reversing the sign of the magnetic interaction energy
${\H}_{12}^M$, without changing ${\cal H}_1$, by applying an
operator $\theta$ on subsystem $1$. For particles with permanent
magnetization fixed relative to the particle, a  rotation of each particle
can reverse the direction of magnetization. Such a rotation keeps the
residual interactions unchanged only if the particle shape has an
axis of 2-fold symmetry perpendicular to its magnetization. Hence
at least one operator $\theta$ exists, and our proof applies, for
systems of identical particles with the required rotational symmetry
in shape.

Some kinds of small particles, including many used in
ferrofluids~\cite{rr}, exhibit superparamagnetism. Dipole moments
rotate by Neel relaxation~\cite{luo}, or possibly quantum
tunneling~\cite{awschalom}, without requiring rotation of the particle
itself. To describe the superparamagnetic classical particles, one includes
in $\Omega_i$, in addition to the Euler angles, a discrete variable 
$O_i=\pm 1$ specifying that the magnetization is parallel $(+1)$ or opposite
$(-1)$ to a direction fixed in the particle, and $\int d\Omega_i$ includes a
sum over $O_i$. The $\theta$ operator is the map $O_i \rightarrow -O_i$,
applied to every particle. For a quantum particle, the corresponding operation
is time reversal applied to the particle's magnetization. In either case,
the $\theta$ operator has the properties specified in 
section~\ref{dipolar_systems}, so the argument given there shows that any 
system of hard core superparamagnetic particles, with any shape of particle,
has a thermodynamic limit.

\subsection{Dipolar systems with central forces}
\label{central}

Consider a system of particles interacting with Hamiltonian
\begin{equation}
\label{model}
\H=\H^M+\H^{(n)}+\H^{central}
\end{equation}
where $\H^M$ is the point dipole interaction~(\ref{HDD}), 
\begin{equation}
\H^{(n)}= \sum_{i<j=1}^N  { A \over {r_{ij}}^n}
\end{equation}
is a repulsive interaction with $A>0$ and $ n>3$, and $\H^{central}$ is any
stable~(\ref{stable}) and tempered~(\ref{tempered}) potential that is
spherically symmetric. Define $\H^{(n)}+\H^{central}$ to be the
residual interaction $\H^R$. The upper bound~(\ref{upper_bound})
follows exactly as in section~\ref{dipolar_systems} because
$\H^{(n)}$ and $\H^{central}$ are both tempered and rotationally
invariant. The proof of a lower bound~(\ref{lower_bound}) for such
systems is more complicated than for hard core particles because there 
is no minimum distance of separation between the point dipoles.

To prove stability~(\ref{stable}) and hence a lower
bound~(\ref{lower_bound}), it suffices to prove that $\H^M+\H^{(n)}$ is
stable, since $\H^{central}$ is stable by assumption. Consider some 
configuration of a
finite system of $N$ particles.  Let $2R_i$ be the distance from the
$i^{th}$ particle to its nearest neighbor. The magnetic interaction
energy remains unchanged if each particle is replaced by a sphere of
radius $R_i$ with uniform magnetization and the same dipole moment
$\boldmu$.  The self energy of such a sphere is $\mu^2/2R_i^3$.  
Since ${ \H^M+(\mu^2/2)\sum_{i=1}^N R_i^{-3}\ge 0}$ by 
positivity of field
energy (see discussion in section~\ref{hard_core_particles}), and
$\H^{(n)} \ge (A/2)\sum_{i=1}^N ({2R_i})^{-n}$, 
we write 
\begin{equation}
\label{inequality}
\H^M+\H^{(n)} \ge - \sum_{i=1}^N {{{\mu}^2} \over {2 {R_i}^3}}
+ \sum_{i=1}^N {A \over {2({2R_i})^{n}}}.
\end{equation}
Define a generalized mean
\begin{equation}
\label{mean}
G(t)=({1 \over N} \sum_{i=1}^N {1 \over {R_i}^t})^{1 \over t }.
\end{equation}
Using the property that $G(t)$ increases monotonically  for positive 
$t$~\cite{as} we find
\begin{equation}
\label{powers}
{1 \over N} \sum_{i=1}^N {1 \over {R_i}^{n}} \ge ({1 \over N}\sum_{i=1}^N
{1 \over {R_i}^{3}})^{n/3},
\end{equation}
for $n>3$. Combining equation~(\ref{powers}) with equation~(\ref{inequality})
we write
\begin{equation}
\label{inequality2}
\H^M+\H^{(n)} \ge -N ({{\mu}^2 \over 2} X - {A \over 2^{n+1}} X^{n/3}),
\end{equation}
where 
$ X\equiv ({1/ N} )\sum_{i=1}^N {R_i}^{-3}$. 
The bound~(\ref{inequality2}) for $\H^M+\H^{(n)}$ has a minimum because $n>3$ 
and $A>0$. In particular
\begin{equation}
\H^M+\H^{(n)} \ge -N [{\mu^2\over 2}({n-3 \over n}) ({3~2^n\mu^2 \over 
{n A}})^{3 \over (n-3)}].
\end{equation}
Our model~(\ref{model}) is therefore stable.

Let's apply this general proof to some special cases. Our proof
applies to generalized Lennard-Jones particles with dipole
interactions.  The Hamiltonian for such a system is
\begin{equation}
\H=\H^M+\H^{LJ},
\end{equation}
where the Lennard-Jones potential is
\begin{equation}
\H^{LJ}=\sum_{i<j} { B \over {r_{ij}}^n}-\sum_{i<j} 
{ C \over {r_{ij}}^m}.
\end{equation} 
Ruelle~\cite{dr} showed that generalized Lennard-Jones potentials
with $B, C>0$ and $n>m>3$ are stable. To demonstrate stability including 
the dipole interaction, divide the repulsive term into two positive pieces,
$B=A+B'$.  Attribute $B'$ to a new (but still stable) Lennard-Jones
potential and use the remainder $A$ to define the repulsive potential
$\H^{(n)}$,
\begin{equation}
\label{LJ}
\H = {\H^M} + {\H^{(n)}} +{\H^{LJ'}}.
\end{equation}
The Hamiltonian in~(\ref{LJ}) is special case of our
model~(\ref{model}). The Stockmayer fluid~\cite{smf} is the case with
$n=12, m=6$, and hence will also have a shape independent
thermodynamic limit. Dipolar soft spheres~\cite{patey} are the
trivial case C=0 and $\H^{central}=0$.

\subsection{Polarizable Particles}
\label{polar}

Consider a system of identical hard core particles that contain
permanent magnetic moments but are further linearly (i.e.
proportionally to the local field) polarizable. The simplest model for
such systems is the dipole-induced-dipole (DID)
model~\cite{wertheim,oxtoby}. The model consists of spherical
particles with a point dipole moment at the center, and  with the
induced polarization an additional point dipole moment of strength
$\alpha {\bf H}$.  This model lacks stability in general.  For
example, with two spherical particles of polarizability $\alpha$ and
hard core radius $R$, stability is lost when
\begin{equation}
{\alpha \over 4R^3} \ge 1.
\end{equation}
The model fails because of its assumed induced point dipole.  The
point dipole applies rigorously only to an infinitesimal volume.
However, the polarizability $\alpha$ necessarily vanishes in the limit
of zero volume, due to the self-induced demagnetizing field. Finite
size particles can have $\alpha \ne 0$. However, the DID model omits
multipole moments due to the spatial variation of fields and
magnetization inside the particles. Higher order multipole interactions
between particles become important when the particles approach each
other~\cite{NewPatey,ZW_emulsion}, and are required for stability.

Consequently, we work with a more physically realistic
model~\cite{oxtoby}. By incorporating the full magnetic interaction
(dipole and higher moments) and the spatial variation of fields within
a particle, our model satisfies stability in general. This model
represents the polarizability of atoms and molecules more accurately
than the DID model. Each particle has a permanent magnetization density 
which is constant in the interior of the particle, but whose direction is 
determined by the orientation of the particle. (For example, imagine
that the particles are prolate ellipsoids with magnetization along the
long axis.) The particles are ``hard'', so that their volumes cannot overlap.
Consequently, the permanent magnetization is a vector field 
${\bf M}^p({\bf r})$, equal to zero unless ${\bf r}$ is inside some particle,
where it takes on a value whose magnitude is independent of the particle but
whose direction is tied to the particle's orientation $\Omega$.

In addition, each particle contains linearly polarizable material giving rise
to an induced magnetization
\begin{equation}
\label{suscep}
{\bf M}^i({\bf r})=\boldchi({\bf r}) \cdot {\bf H}({\bf r}),
\end{equation}
where ${\bf H}({\bf r})$ is the total magnetic field at ${\bf r}$, and
the susceptibility tensor $\boldchi({\bf r})$ is zero unless ${\bf r}$ is
inside some particle, where its value is independent of ${\bf r}$ but tied
to the orientation of the particle: that is the principal values of 
$\boldchi$ and the relationship of the principal axes of $\boldchi$ to the
orientation of the particle is same for every particle.

The total magnetic field ${\bf H}({\bf r})$ in~(\ref{suscep}) is
\begin{equation}
\label{field}
{\bf H}({\bf r})={\bf H}^p({\bf r})+{\bf H}^i({\bf r}),
\end{equation}
where ${\bf H}^p({\bf r})$ is the field from permanent magnetization density
${\bf M}^p$, and is given by~(\ref{demag-field}) with  ${\bf M}$ set equal
to ${\bf M}^p$, whereas ${\bf H}^i({\bf r})$ is due to the induced 
magnetization: in~(\ref{demag-field}) set ${\bf M}$ equal to ${\bf M}^i$.
Note that even an isolated particle will have an induced magnetization because
the demagnetizing effect will give rise to a non-zero ${\bf H}^p$, and
the total ${\bf H}$ inside the particle must be determined self-consistently,
as it both induces a magnetization, (\ref{suscep}), and is partly 
(${\bf H}^i$) determined by that magnetization.

Because of this requirement of consistency between total field and
magnetization, the magnetic interaction of polarizable particles is a
multi-body interaction, and is much more complicated than the pairwise
multipole interactions of permanently magnetized particles discussed
in section~\ref{hard_core_particles}. To write down the interaction energy
$\H^M$ of a configuration of polarizable particles, it is convenient
to calculate the work done assembling the configuration, starting with
the particles well separated from each other at infinity. For any
arrangement of particles, the total magnetic energy is
\begin{eqnarray}
\label{pol-nrg}
\H^T ={1\over{8 \pi}} \int d^3 
{\bf r}~|{\bf H}^p|^2 -{1\over{2}} \int d^3 {\bf r}
~{\bf H}^p \cdot {\bf M}^i.
\end{eqnarray}
The first term is the demagnetization energy~(\ref{energy}) of the
permanent magnetization distribution, and the second term represents the 
work done in introducing linearly polarizable material into this permanent
field. Evaluating~(\ref{pol-nrg}) for an isolated particle defines
the self energy $E^{self}$ per particle. The difference
\begin{equation}
\label{define}
\label{mag-int}
\H^M \equiv \H^T - N E^{self}
\end{equation}
equals the work done bringing initially isolated polarizable particles
into interaction with each other.

The stability~(\ref{stable}) and hence the lower
bound~(\ref{lower_bound}) for this system follows from the positivity
of field energy.  Rewrite the magnetic energy in
equation~(\ref{pol-nrg}) as (see appendix A for details)
\begin{equation}
\label{pol-nrg2}
\H^T= {1\over{8 \pi}} \int d^3 
{\bf r}~{\bf H}\cdot ({\bf 1}+4 \pi\boldchi)\cdot{\bf H}.
\end{equation}
Since ${\bf 1}+4\pi \boldchi$ is positive definite,  $\H^T \ge 0$, hence
\begin{equation}
\H^M \ge - N E^{self}.
\end{equation}
The stability and lower bound then follow as in the case of identical
hard core particles.

To prove the upper bound~(\ref{upper_bound}), consider two 
subsystems $1$ and $2$ separated by a distance $d \ge d_0 > 0$ as in 
section~\ref{formal}. Write the interaction Hamiltonian between these two 
subsystems as
\begin{equation}
\label{int-nrg}
{\H_{12}^M}=\H^M-\H_1^M-\H_2^M,
\end{equation}
where $\H_1^M$ and $\H_2^M$ are the magnetic interaction Hamiltonians
for the subsystems by themselves; i.e. $\H_1^M$ is the energy of
subsystem $1$ with subsystem $2$ placed infinitely far away. We show in
appendix B that, for positive-definite $\boldchi$,
\begin{equation}
\label{even-odd}
{\H_{12}^M}= {\cal O}_{12}^M+{\cal N}_{12}^M,
\end{equation}
where ${\cal O}_{12}^M$ is odd under reversal of the permanent
magnetization of particles in subsystem $1$ by the $\theta$ operator,
and ${\cal N}_{12}^M \le 0$. Since ${\cal O}_{12}^M$ is odd, its
ensemble average vanishes. Then
\begin{equation} 
\langle \H_{12}^M \rangle_{\lambda=0}=
\langle {\cal N}_{12}^M\rangle_{\lambda=0} \le 0,
\end{equation}
which is sufficient to prove the upper bound on $F$, as discussed in
section~\ref{dipolar_systems}.

\subsection{Quantum Systems}
\label{quantum}

Our proofs extend to quantum systems. Consider a system of $N$
identical spin $S$ particles in volume $V$, which may obey Boltzmann,
Fermi-Dirac or Bose-Einstein statistics. The Hamiltonian is
\begin{equation}
\label{qm}
\H= {\cal K}+\H^M+\H^{EX}+\H^R.
\end{equation}
Here ${\cal K}= \sum_{i=1}^N p_i^2 / 2m$ is the kinetic energy
operator.  The exchange interaction~\cite{ashcroft} is
$\H^{EX}={{1}\over{2}} \sum_{ij} J(r_{ij}) {\bf S}_i\cdot{\bf S}_j$,
where ${\bf S}_i$ is the spin operator for particle $i$, $r_{ij}$ is the 
distance between particles $i$ and $j$, the couplings 
$J(r_{ij})$ are assumed to satisfy conditions consistent with stability, and
the sum is over all $i\ne j$. The dipole 
interaction  $\H^M$ is given by~(\ref{HDD}) with dipole moments 
$\boldmu_i = g{\bf S}_i$.  The residual interaction $\H^R$ may be any  
stable and tempered interaction that remains unchanged under simultaneous spin 
reversal of all particles, such as the hard core or central force interactions 
discussed in sections \ref{hard_core_particles} and \ref{central}. Implicit 
in our definition of the Hamiltonian is the confinement of particles in a 
region $\R$ of volume $V$ with hard wall boundary conditions. The 
partition function is
\begin{equation}
\label{partfunc}
Z={{1}\over{C}} ~{\rm Tr}~{\big \{}e^{-\H /k_B T}{\big \}}
\end{equation} 
where the trace {\rm Tr} is carried out over states of appropriate 
symmetry with 
respect to interchange of particles. For Boltzmann particles  $C=N!$ and 
for fermions and bosons $C=1$. The free energy $F$ is $-k_B T \log Z$. 

The stability of $\H$, in the sense that its spectrum has a lower bound
proportional to $N$, see~(\ref{stable}), follows from the positivity of
${\cal K}$ (no negative eigenvalues) and the stability of $\H^M+\H^{EX}+\H^R$.
Lower bounds on classical energies of the sort derived in 
sections \ref{hard_core_particles} and \ref{central} are easily 
extended to operator inequalities which
demonstrate the stability of $\H^M+\H^R$. That this stability is preserved
upon adding $\H^{EX}$ requires a suitable choice of $J(r)$. For example, in the
case of hard core particles it suffices that $J(r)$ be bounded and decrease
more rapidly than $r^{-3-\epsilon}$ for some $\epsilon>0$.

We now address the upper bound on the free energy. Write the Hamiltonian  in 
the form  
\begin{equation}
\label{confined}
\H=\H_1 + \H_2+ \H_{12}
\end{equation}
where $\H_1$ includes all terms in~(\ref{qm}) (one particle, two-particle,
etc.) involving only particles in the set $\S_1$ with 
labels $1, 2,..., N_1,$ $\H_2$ includes the terms involving only particles in 
the set $\S_2$ with labels $N_1+1, N_1+2,..., N_1+N_2,$ and 
$\H_{12}$ all the remaining terms. The interaction energy $\H_{12}$ is a sum
of magnetic dipole, exchange, and residual terms
\begin{equation}
\label{int}
\H_{12}=H_{12}^M+\H_{12}^{EX}+\H_{12}^R.
\end{equation}
 It will also be convenient
to define a scaled Hamiltonian
\begin{equation}
\label{scaled}
\H(\lambda)=\H_1 + \H_2+ \lambda \H_{12},~0\le \lambda \le 1.
\end{equation}

To begin with, we assume that  the $N_1$ particles with labels in 
$\S_1$ are confined to a region of space $\R_1 \subset \R$  with  
volume $V_1$, while the $N_2=N-N_1$ particles with labels in $\S_2$ 
are confined to a  region $\R_2 \subset \R$ of volume $V_2$, 
and that the minimum separation of $\R_1$ and $\R_2$ is
a distance $d \ge d_0$. Note that (in general) $\R$ is larger than 
$\R_1 \cup \R_2$. The Hilbert space of this system is spanned by 
product states of the form
\begin{equation}
\label{unsym}
|\psi_{n,m}^\U\rangle= |\phi_m(1,...,N_1) \rangle
~|\chi_n(N_1+1,...,N_1+N_2) \rangle.
\end{equation}
where the integer arguments are particle labels. The $\{|\phi_m\rangle\}$ form
a complete set of states for the $N_1$ particles in $\R_1$ with appropriate
symmetry under interchange of particles and $\{|\chi_n\rangle\}$ is a similar 
set for the $N_2$ particles in $\R_2$.  The product states~(\ref{unsym}) have 
no symmetry for the interchange of particles between the two sets $\S_1$ and 
$\S_2$, as indicated by the superscript $\U$. Due to this lack of symmetry the 
partition function
is given by
\begin{equation}
\label{unsym-partf}
Z_{N_1,N_2}^{\C,\U}={1 \over D} {\rm Tr} ~\{~e^{-\H/k_BT}\}
\end{equation}
where $D=N_1!N_2!$ for Boltzmann particles and $D=1$ for fermions and
bosons. If this partition function is evaluated using $\H$ with $\lambda=0$
in~(\ref{scaled}), $-k_BT$ times its logarithm is the sum $F_1+F_2$ 
of free energies for the separate systems of particles in regions $\R_1$ 
and $\R_2$, each evaluated as if the other region did not exist, since 
the interactions between the two sets of particles have been ``turned off''.

An upper bound on the free energy $F$ of the full system of $N$ particles
confined to region $\R$, in the form of the equation~(\ref{upper_bound}) is 
obtained through the following steps.

1) ``Turn on'' the interaction between the particles in regions $\R_1$ and 
$\R_2$  by letting $\lambda$ increase from $0$ to $1$. The resulting
free energy is denoted by $F_{N_1,N_2}^{\C,\U}$. The superscript 
$\C$ indicates that particles are confined to regions $\R_1$ and $\R_2$.

2) Remove the constraint that only particles in set $\S_1$ 
are found in $\R_1$ and only those in set $\S_2$ are 
found in $\R_2$. Any particle
may be anywhere in the union of $\R_1$ and $\R_2$, provided there are exactly
$N_1$ particles in $\R_1$ and $N_2$ particles in $\R_2$. This involves
introducing appropriately symmetrized wavefunctions. 

3) Relax the constraint that precisely $N_1$ particles are in $R_1$ and $N_2$
in $\R_2$. We still require that there be a total of $N$
particles in the union of $\R_1$ and $\R_2$.

4) Relax the constraint that the particles lie in either $\R_1$ or $\R_2$ , 
so that all particles can be anywhere in the larger region $\R$.

Since steps 2, 3, and 4 do not require special attention to long range
interactions, they are discussed in Appendix C. Basically, each time
a constraint is relaxed the free energy decreases, except for step 2, where
it remains constant. Thus the upper bound obtained for 
$F_{N_1,N_2}^{\C,\U}$ applies to $F$.

For step 1 we use the fact that $F_{N_1,N_2}^{\C,\U}$ is a concave function
of $\lambda$, and therefore
\begin{equation}
\label{bogo1}
F_{N_1,N_2}^{\C,\U}(1)\le F_{N_1,N_2}^{\C,\U}(0)
+(dF_{N_1,N_2}^{\C,\U}/d\lambda)_{\lambda=0},
\end{equation}
which is known as the Bogoliubov inequality~\cite{falk}. The first term on 
the right is
$F_1+F_2$, and the second is
\begin{equation}
\label{average}
\langle \H_{12}\rangle_{\lambda=0}=\langle \H_{12}^M\rangle_{\lambda=0}+
\langle \H_{12}^{EX}\rangle_{\lambda=0}+\langle \H_{12}^R \rangle_{\lambda=0},
\end{equation}
where the averages are with respect to Boltzmann weights with $\lambda=0$. 
For example
\begin{equation}
\label{trace2}
\langle \H_{12}^M\rangle_{\lambda=0}={\rm Tr}~\{\H_{12}^M~e^{-(\H_1+\H_2)
/k_BT}~\}
/{\rm Tr}~\{e^{-(\H_1+\H_2)/k_BT}~\}.
\end{equation}
The third term in~(\ref{average}) is bounded by the upper bound $\Delta_{12}$
on $\H_{12}^R$. The first and second terms vanish for the following reason.
Let $\theta$ be
the anti-unitary spin reversal operator which reverses all the spins of the 
particles in collection $\S_1$. It is a symmetry of $\H_1$, because $\H_1^M$
and $\H_1^{EX}$ involve products of two spin operators and $\H_1^R$ is, by 
assumption, invariant under the reversal of all spins. Consequently, if the
$\{|\phi_m \rangle\}$ are the eigenstates of $\H_1$, the states 
$\theta |\phi_m \rangle=|\hat{\phi}_m \rangle$ are also eigenstates with the
same eigenvalues. In evaluating the trace in the numerator of~(\ref{trace2})
we can employ a complete set $\{|\phi_m \rangle|\chi_n \rangle\}$, where the
$\{|\chi_n \rangle\}$ are the eigenstates of $\H_2$, or, equivalently,
$\{|{\hat\phi}_m \rangle|\chi_n \rangle\}$. However, 
\begin{equation}
\label{commutes}
\langle \hat{\phi}_m| \langle{\chi}_n|\H_{12}^M|\hat{\phi}_m \rangle |\chi_n 
\rangle=-\langle \phi_m|\langle{\chi}_n| \H_{12}^M |\phi_m \rangle|\chi_n 
\rangle 
\end{equation}
because
\begin{equation}
\label{anticommutes}
\langle \hat{\phi}_m| {\bf S}_i |\hat{\phi}_m \rangle = 
-\langle \phi_m| {\bf S}_i |\phi_m \rangle
\end{equation}
for $1\le i \le N_1$, and $\H_{12}^M$ is a sum of pairwise products of spin 
operators, one from the collection $\S_1$ and one from 
$\S_2$. Thus 
$\langle \H_{12}^M\rangle_{\lambda=0}$ is equal to its negative and vanishes.
The same argument applies to $\langle\H_{12}^{EX}
\rangle_{\lambda=0}$. 

Consequently, after following the remaining steps 2, 3, and 4 contained in
appendix C, we have a bound corresponding to~(\ref{upper_bound}) in the 
classical case. This bound together with that in~(\ref{lower_bound}),
which (as already noted) follows from the stability of $\H$, completes the 
proof of the existence of a thermodynamic limit in the quantum case, see
section~\ref{formal}.

It is possible to treat the translation degrees of freedom classically
and the spin quantum mechanically. The proof of a thermodynamic limit 
for such a ``semi quantum''
model is similar to the ``fully quantum'' treatment. The averages
of $\H_{12}^M$ and $\H_{12}^{EX}$ involve sums over all spin states and
integrals over all particle positions. Using the spin reversal operator
$\theta$, introduced earlier in this section, and doing the sums over all spin
states first,  one sees that the average values of $\H_{12}^M$ and 
$\H_{12}^{EX}$ vanish. The residual interactions $\H_{12}^R$ are bounded by 
$\Delta_{12}$, and therefore our proof goes through.

One may define another ``semi quantum'' model of classical spins and 
classical dipoles with center-of-mass motion of the particles treated 
quantum mechanically. The averages of
$\H_{12}^M$ and $\H_{12}^{EX}$ contain integrals over all spin
orientations and sums over all spatial wavefunctions in this case.
The proof is similar to that of classical particles. Doing the
integrals over particle spins first, gives zero for the average of
$\H_{12}^M+\H_{12}^{EX}$, because they are both odd with respect to
spin reversal. The sum over the spatial wavefunctions then yields the
desired upper bound on $\langle \H_{12} \rangle_{\lambda=0}$  and 
our proof goes through. We can even include classical polarizability 
(section~\ref{polar}) in such a model. In this case the average of $\H_{12}^M$
is non positive, and again we get the required upper bound~(\ref{upper_bound})
on the free energy.

\section{Electric Polarization}
\label{electric}

So far we phrased our discussion entirely in terms of magnetic
interactions. However our proof applies to many electrically polarized
or polarizable materials as well. Electric fields ${\bf E}$ and 
${\bf D}$ fields satisfy the same Maxwell equations as magnetic fields 
${\bf H}$ and ${\bf B}$, respectively, provided no free charges and 
currents are present. By replacing fields ${\bf H}$, ${\bf B}$ and 
magnetization ${\bf M}$ with fields ${\bf E}$, ${\bf D}$ and polarization 
${\bf P}$, respectively, our proofs run exactly as for magnetic materials. 
We assume stability, tempering of the residual interactions, and the
existence of a $\theta$ operator. Thus we prove the existence of a
shape independent thermodynamic limit for the electric analogue of
each of the classical models discussed in sections~\ref{models}.

For ferroelectric identical hard core particles, use the Hamiltonian
\begin{equation}
\H= \H^P+\H^{HC}
\end{equation}
where $\H^P$ is the electric interaction between particles. The proof follows
section~\ref{hard_core_particles} with the replacement of fields discussed 
at the beginning of this section.  For electric dipolar systems with
central forces the interaction Hamiltonian is
\begin{equation}
\H=\H^P+\H^{(n)}+\H^{central}.
\end{equation}
The proof goes through as in section~\ref{central}, with the existence
of a shape independent thermodynamic limit for electric dipolar soft
spheres and Stockmayer fluids as special cases.  For
electrically polarizable particles, $\H=\H^{P} + \H^{HC}$ where
$\H^P$, the electric interaction between polarizable particles, is
defined analogously with $\H^M$ in equation~(\ref{mag-int}). After the
replacements of fields and magnetization discussed at the beginning of
this section, the proof proceeds as in section~\ref{polar}. Quantum
models can be treated as discussed in section~\ref{quantum}.

Adding another layer of complexity, we may  define models
combining features of electric and magnetic models already discussed.
The possible variations are too numerous to describe individually. We
simply note here that stability including only magnetic or only
electric interactions ensures stability with the two added together.
Finding a $\theta$ operator may be more difficult. For example, a
$180^{\circ}$ rotation axis must lie perpendicular to both ${\bf M}$
and ${\bf P}$.

For the crude dipolar particle models discussed above, the electric
and magnetic problems are fully equivalent.  For applications to realistic
models of specific materials, however, it is generally harder to find 
a $\theta$ operator in the case of electric materials. The analogue of time
reversal (which provides a $\theta$ operator for superparamagnetic
particles) is charge conjugation. This has the undesirable effect of
turning matter into antimatter! Without a $\theta$ operator we cannot
apply our proof.

For an example of a model outside our proof, consider a fluid of H$_2$O 
molecules (water). Modeled as a dipolar hard sphere or a Stockmayer fluid 
the thermodynamic limit follows from our discussions above using rotation 
as the $\theta$ operator. Real H$_2$O molecules lack this rotational symmetry. 
For example, the TIPS 3 site model of water~\cite{tip} places positive 
charges on the hydrogens and a negative charge on oxygen. Coulomb
interactions between all intermolecular pairs of charges, and a  
Lennard-Jones interaction between the oxygen atoms, gives the interaction
\begin{equation}
\label{water}
\H_{mn}= \sum_{ij} {q_i q_j \over r_{ij}}
+{A \over r_{OO}^{12}}-{C \over r_{OO}^6},
\end{equation}
between any two water molecules $m,n$, where $i$ and $j$ run over the charges $q_i$ and $q_j$, respectively,
on molecule $m$ and $n$, $r_{ij}$ is the separation between charges
and $r_{OO}$ is the separation between oxygen ions.  For the above
model there is no symmetry operation available which can reverse all
the electric interactions while keeping the residual (Lennard-Jones)
interactions unchanged. A $\theta$ operator therefore does not exist
for this model and we cannot apply our proof.  We suppose that our
qualitative argument (section \ref{nofield}) demonstrating existence
of a shape independent thermodynamic limit based on domain formation
still holds, but for technical reasons we cannot prove it.

Yet another example is provided by models of ferroelectric materials
with mobile charges. Charges transfer among dissimilar chemical
elements, so no $\theta$ operator is likely to exist. Furthermore, the
definition of polarization becomes ambiguous, depending on how the
unit cell is defined in the case of a crystal~\cite{Vanderbilt}, or on
the surface charge for non-crystalline materials. The question of
appropriate thermodynamic limits for model ferroelectrics with mobile
charges remains under discussion~\cite{Gonze}.

For models containing only bound charges, our qualitative argument
(section~\ref{nofield}) suggests the equilibrium state has no
depolarizing energy density. When the microscopic coulomb charges are 
taken into account, including the possibility of molecular dissociation 
and the Fermi statistics of the constituent particles, the problem falls
into the class of materials for which Lieb~\cite{lieb} proved a
thermodynamic limit. Free electric charges screen the $1/r$ Coulomb
potential, restoring the thermodynamic limit.

\section{Conclusions}
\label{conclusion}

	We proved the existence and shape independence of the free energy
density for a variety of dipolar systems. Three essential conditions were
identified: stability, tempering of residual interactions, and the
existence of a $\theta$ operator that commutes with ${\H}_1$, $\H_2$ and
$\H_{12}^R$ while reversing the sign of $\H_{12}^M$. Our proof covers 
systems of identical hard core particles with
uniform permanent magnetization. We also treat dipolar soft spheres 
and Stockmayer fluids, systems with magnetizable or polarizable material, 
and we consider electric as well as magnetic dipoles. Except for the case of
super-paramagnetic particles, the existence of a $\theta$ operator requires
symmetries such as a 2-fold axis of rotational symmetry perpendicular
to the magnetization/polarization of each particle.

	Having proven shape independence of the free energy we now
consider some implications of the proof. For the systems covered by our
proof, thermodynamic states and phase diagrams do not depend on size or
shape~\cite{GD}. Intrinsic thermodynamic quantities such as pressure and
chemical potential are independent of sample shape and position within a 
sample. When calculating free energies of magnetized states, care must be
taken to remove the depolarizing field if uniform magnetization is
assumed.  Failure to do so leads to either boundary-condition or 
shape dependence of thermodynamic properties~\cite{Gonze,GD,LT,petschek}.  
Two convenient ways to remove the demagnetizing fields are to study 
highly prolate ellipsoids or to use tin-foil boundary 
conditions~\cite{leeuw}.

	Computer simulations~\cite{PateyWeis} suggest dipolar
fluids such as ferrofluids may spontaneously magnetize. Experiments on
supercooled CoPd alloys~\cite{Platzek} claim to observe a
spontaneously magnetized metastable state.  How then can we reconcile
shape independence of the free energy with these reports of
spontaneously polarized liquid states? Domains
or textures must form, as in equilibrium solid ferromagnets.
Liquids lack the crystalline anisotropy required to
sustain a sharp domain wall (see Fig. 1b).  De Gennes and Pincus~\cite{DP} 
note, however, that domain wall thickness should be comparable to system 
dimensions in a magnetic liquid. We conclude that a
spontaneously polarized liquid has a position dependent axis of
polarization that rotates so as to lie tangent to the sample surface.  
Since vector fields tangent to the surface of 
simply connected volumes exhibit singularities (the ``hairy
billiard ball'' problem), magnetized fluid droplets must contain defects in
the magnetization field ${\bf M}$. Possible textures include line (Fig. 1c)
or point defects (not shown). Away from such defects the magnitude of
magnetization $|{\bf M}|$ is independent of position within the sample.  
In computer simulations a uniformly polarized state is observed because 
the Ewald summations~\cite{leeuw} drop the surface pole energy 
(via ``tin-foil'' boundaries) and thus mimic an infinite, boundary-free,
medium.

	Our proof is valid only in zero applied field. The case of an
applied field is still open. The free energy in a field depends on shape as
outlined in section~\ref{demagfield}. The free energy increase due to 
the demagnetizing field causes a droplet of paramagnetic liquid to 
elongate in an external field, minimizing its magnetic energy by 
reducing $D$~\cite{elong}. However, we conjecture the existence
of an intrinsic thermodynamic energy, defined in
equation~(\ref{conjecture2}) by subtracting the shape dependent
demagnetizing energy from the full shape dependent free energy. 

	Finally, let's consider some dipolar systems to which we are
unable to apply our proof due to the lack of a $\theta$ operator.
Particles with non-symmetric shapes and magnetization fixed with
respect their body (Fig.~\ref{f2}a) provide an example. Each particle
is a cube with protrusions (conical, hemispherical and cubic) on three
faces and matching indentations on the opposite three faces. Each face
of one particle fits exactly into the corresponding opposite face of
another particle. Recall that we exploit symmetries of the Hamiltonian
when applying a $\theta$ operator to the internal coordinates
$\Omega_i$. The $\theta$ operator, if it exists, reverses the sign of
$\H_{12}^M$ while leaving $\H_1$ invariant. Rotations are not a
symmetry of the internal coordinates for these particles. There is
no evident symmetry of the Hamiltonian which could be used
as a $\theta$ operator.

When these particles are tightly packed (i.e. the limit of infinite 
pressure and hard cores), they align parallel to each other. ${\bf M}$ 
is uniform throughout space and therefore the  
thermodynamic limit does not exist. At finite pressure it is  possible 
to accumulate enough free volume to insert domain walls, (Fig.~\ref{f2}b), 
so we expect that the thermodynamic limit does 
exist even though the conditions of our proof are not obeyed. That is, our 
conditions are sufficient but not in all cases necessary.

	Another interesting example is provided by hydrogen absorption
in metals~\cite{wagner}. Interstitial hydrogen creates elastic strain
fields that fall off as $1/r^3$, and hence are referred to as dipoles.
This is a rather confusing notation, since the angular dependence is
actually a sum of a quadrupole and a monopole contribution. The
monopole term is the trace of the strain tensor and governs lattice
dilation. Provided the system remains coherent (the lattice structure
is dislocation free), the monopole interactions
are shape dependent, attractive and infinite ranged. The
interactions reduce the energy but create no force. In the coherent
state, the phase diagram for liquid-gas transitions of hydrogen in a metal
depends on the shape of the metal~\cite{experiment}.

	The coherent state itself is a metastable state. The true
equilibrium state is incoherent, with dislocations relaxing the
lattice strain. If one allows for dislocations, the elastic interactions 
become short ranged and
the thermodynamic limit is restored. The sample itself, however, may
have disintegrated into a fine powder!

\acknowledgements

We acknowledge useful discussions with H. Zhang, Ralf Petscheck, JJ.
Weis and D. Levesque. This work was supported by NSF grant DMR-9221596
and by the A.P. Sloan foundation.

\appendix
\section{Total magnetic energy for polarizable dipoles}
\label{appa}

We start with equation~(\ref{pol-nrg2}) for $\H^T$ and show that it
equals equation~(\ref{pol-nrg}). Use equations~(\ref{suscep}) for
induced magnetization ${\bf M}^i$ and (\ref{field}) defining the
permanent and induced fields ${\bf H}^p$ and ${\bf H}^i$ to rewrite
equation~(\ref{pol-nrg2}) as
\begin{eqnarray}
\label{pol-nrg3}
\H^T= {1\over{8 \pi}} \int d^3{\bf r}~|{\bf H}^p|^2+
{1\over{8 \pi}} \int d^3{\bf r}~|{\bf H}^i|^2 +{1\over{4 \pi}}
\int d^3{\bf r}~{\bf H}^p \cdot
{\bf H}^i
\\ \nonumber
 + 
{1\over{2}} \int d^3{\bf r}~{\bf H}^p \cdot{\bf M}^i 
+ {1\over{2}} \int d^3{\bf r}~{\bf H}^i \cdot{\bf M}^i
\end{eqnarray}
For any two magnetization distributions ${\bf M}_1({\bf r})$, and 
${\bf M}_2({\bf r})$ and the fields ${\bf H}_1({\bf r})$ and 
${\bf H}_2({\bf r})$ caused by them respectively, the following identity
holds~\cite{wfb}:
\begin{eqnarray}
\label{iden}
{1\over{8 \pi}} \int d^3{\bf r}~{\bf H}_1\cdot{\bf H}_2
=-{1\over{2}} \int d^3{\bf r}~{\bf H}_1\cdot{\bf M}_2
=-{1\over{2}} \int d^3{\bf r}~{\bf H}_2\cdot{\bf M}_1
\end{eqnarray}
Since equation~(\ref{iden}) holds for any two arbitrary magnetization
distributions, we set ${\bf M}_1={\bf M}_2$ equal to the 
induced magnetization ${\bf M}^i$. Then equation~(\ref{iden}) gives
\begin{equation}
\label{ind}
{1\over{8 \pi}} \int d^3{\bf r}~|{\bf H}^i|^2
=-{1\over{2}} \int d^3{\bf r}~{\bf H}^i \cdot {\bf M}^i.
\end{equation}
Similarly, setting ${\bf M}_1={\bf M}^p$ and ${\bf M}_2={\bf M}^i$ in
equation~(\ref{iden}) gives
\begin{equation}
\label{perm}
{1\over{8 \pi}}\int d^3{\bf r}~{\bf H}^p \cdot{\bf H}^i
=-{1\over{2}} \int d^3{\bf r}~{\bf H}^p \cdot{\bf M}^i.
\end{equation}
Using equations~(\ref{ind}) and (\ref{perm}) to simplify
equation~(\ref{pol-nrg3}) gives equation~(\ref{pol-nrg}), proving
equality of our expressions~(\ref{pol-nrg}) and~(\ref{pol-nrg2}) for
$\H^T$.

\section{Interaction energy between two subsystems of polarizable
dipoles}
\label{appb}

Let ${\bf H}_1^p({\bf r})$ and ${\bf H}_2^p({\bf r})$ be the fields
due to the permanent polarization in subsystems $1$ and $2$ located
in non-overlapping regions $\R_1$ and $\R_2$. The induced magnetizations
in the two subsystems can be written in the form
\begin{eqnarray}
\label{m-ind1}
{\bf M}_1^i({\bf r})={\bf M}_1^s({\bf r})+{\bf M}_1^{'}({\bf r}), \\ \nonumber
{\bf M}_2^i({\bf r})={\bf M}_2^s({\bf r})+{\bf M}_2^{'}({\bf r}),
\end{eqnarray}
where ${\bf M}_1^s({\bf r})$ is the induced magnetization which would be 
present in subsystem $1$ were subsystem $2$ absent, and ${\bf M}_2^s({\bf r})$
that of subsystem $2$ were subsystem $1$ absent.

Using equations~({\ref{pol-nrg}) for $\H^T$ and ({\ref{mag-int}) for
$\H^M$ to find the interaction Hamiltonians $\H_1^M$, $\H_2^M$ and $\H^M$
for the two subsystems and the whole system, respectively, we write the
interaction energy in~(\ref{int-nrg}) as
\begin{eqnarray}
\label{int-nrg2}
{\H_{12}^M}={1\over{4 \pi}} \int d^3 {\bf r}~{\bf H}_1^p
\cdot {\bf H}_2^p 
-{1\over{2}} \int d^3{\bf r}~{\bf H}_2^p \cdot {\bf M}_1^s
-{1\over{2}} \int d^3{\bf r}~{\bf H}_1^p \cdot {\bf M}_2^s \\ \nonumber  
-{1\over{2}} \int d^3{\bf r}~({\bf H}_1^p+{\bf H}_2^p) \cdot {\bf M}_1^{'} 
-{1\over{2}} \int d^3 {\bf r}~({\bf H}_1^p+{\bf H}_2^p)
\cdot {\bf M}_2^{'}.
\end{eqnarray}
Break ${\H_{12}^M}$ into odd and non-positive components 
${\H_{12}^M}= {\cal O}_{12}^M+{\cal N}_{12}^M$, where
\begin{equation}
\label{odd}
{\cal O}_{12}^M={1\over{4 \pi}} \int d^3 {\bf r}~{\bf H}_1^p
\cdot {\bf H}_2^p \\ \nonumber
-{1\over{2}} \int d^3{\bf r}~{\bf H}_2^p \cdot {\bf M}_1^s 
-{1\over{2}} \int d^3{\bf r}~{\bf H}_1^p \cdot {\bf M}_2^s 
\end{equation}
is odd under reversal of the permanent magnetization of particles in 
subsystem $1$ by the $\theta$ operator, and
\begin{equation}
\label{even}
{\cal N}_{12}^M=-{1\over{2}} \int d^3{\bf r}~({\bf H}_1^p+{\bf H}_2^p)
\cdot {\bf M}_1^{'}-{1\over{2}} \int d^3 {\bf r}~({\bf H}_1^p+{\bf H}_2^p)
\cdot {\bf M}_2^{'}
\end{equation} 
is non-positive as we now show.  A theorem by Brown~\cite{wfb} states
that for a paramagnetic polarizable material in an applied field, the
unique induced magnetization ${\bf M}^i$ given by
equation~(\ref{suscep}) minimizes the total magnetic energy. Applying
that theorem to our system we observe that
\begin{eqnarray}
\label{ineq}
 -{1\over{2}} \int d^3{\bf r}~({\bf H}_1^p+{\bf H}_2^p) \cdot 
{\bf M}_1^i
-{1\over{2}} \int d^3{\bf r}~({\bf H}_1^p+{\bf H}_2^p) \cdot {\bf M}_2^i
\le \\ \nonumber
-{1\over{2}} \int d^3{\bf r}~({\bf H}_1^p+{\bf H}_2^p) \cdot {\bf M}_1^s
-{1\over{2}} \int d^3{\bf r}~({\bf H}_1^p+{\bf H}_2^p) \cdot {\bf M}_2^s.
\end{eqnarray}
because the interaction-induced magnetization ${\bf M}^i$ has lower
energy than the isolated self-magnetization ${\bf M}^s$.  Upon replacing 
${\bf M}_j^{'}$ in~(\ref{even}), with ${\bf M}_j^i-{\bf M}_j^s$, 
(\ref{m-ind1}), one sees that (\ref{ineq}) implies that
\begin{equation}
{\cal N}_{12}^M \le 0.
\end{equation}

\section{Quantum Systems}
\label{appc}

Here are the details for steps 2, 3, and 4 in section~\ref{quantum}. For
step 2 note that the formal Hamiltonian $\H(\lambda=1)$, defined in 
equation~(\ref{scaled}), is
symmetrical under the interchange of any two particles, since the right
side of~(\ref{confined}) is simply a way of segregating terms in the sum
representing $\H$. To allow any particle to be anywhere
in $\R_1 \cup \R_2$, subject to the requirement of $N_1$ particles in $\R_1$
and $N_2$ in $\R_2$, replace the Hilbert space of the 
form~(\ref{unsym}) with another spanned by states with appropriate symmetries
under interchanging of any pair of particles. We now construct such a
Hilbert space for each type of statistics as indicated by Fisher~\cite{mef}.

First consider identical particles obeying Boltzmann statistics and let
$\{|\mu^j\rangle\}$, j=1,2... be a complete orthonormal set of 
single particle states (including spin) for a particle confined to $\R_1$.
A basis $\{|\phi_m\rangle\}$ for the particles with labels in $\S_1$ can then
be written in the form
\begin{equation}
\label{basis}
|\phi_m(1,...,N_1) \rangle =|\mu^{m_1}(1)\rangle |\mu^{m_2}(2)\rangle
|\mu^{m_3}(3)\rangle...
\end{equation}
where $m$ stands for the sequence $(m_1,m_2...m_{N_1})$ of integer labels.
In the same way, a basis $\{|\chi_n\rangle\}$ for particles with labels in
$\S_2$ can be constructed using single particle states $\{|\nu^k\rangle\}$,
k=1,2... for a particle confined to $\R_2$.

To construct a basis in which any $N_1$ particles are in $\R_1$
and any $N_2$ particles in $\R_2$, we proceed as follows. Consider the
collection $\P$ of permutations $p$ of the integers  $(1,2...N)$, where
$p(j)$ is the image of $j$ under $p$, with the property that
\begin{eqnarray}
p(1)<p(2)<...<p(N_1), \\ \nonumber
p(N_1+1)<p(N_1+2)<...<p(N_1+N_2).
\end{eqnarray}
It is clear that $\P$ contains $P=N!/(N_1! N_2!)$ permutations, one for each 
way of partitioning the integers from $1$ to $N$ into two collections, one
containing $N_1$ and the other containing $N_2$ integers. Then define states
\begin{equation}
\label{boltzmann}
|\psi_{m,n,p}\rangle= \sigma_p |\phi_m(1,...,N_1) \rangle
~|\chi_n(N_1+1,...,N_1+N_2) \rangle,
\end{equation}
where the  operator $\sigma_p$ applies permutation $p$ to the 
$N_1+N_2$ arguments. The set  $\{|\psi_{m,n,p}\rangle\}$  for $m$ and
$n$ defined previously, and $p$ belonging to $\P$, forms an orthonormal basis 
for the $N$ particles in $\R_1 \cup \R_2$, allowing any particle to be in 
either region, subject to the constraint of $N_1$ particles in $\R_1$ and 
the remaining $N_2$ in $\R_2$. 

To prove orthonormality,
\begin{equation}
\label{inner-product}
\langle\psi_{m,n,p}|\psi_{m',n',p'}\rangle=\delta_{m m'} \delta_{n n'} 
\delta_{p p'}, 
\end{equation}
note that factors $\delta_{m m'}, \delta_{n n'}$ follow from orthonormality 
of the single particle states in~(\ref{basis}). To get the factor 
$\delta_{p p'}$ consider $p \ne p'$ both belonging to $\P$. There is at 
least one particle $l$ which is in $\R_1$ under $p$, and in $\R_2$ under $p'$. 
The inner product~(\ref{inner-product}) contains a factor 
$\langle\mu^{m_l}(p(l))|\nu^{n_l}(p'(l))\rangle$ that vanishes 
because the $|\mu^j\rangle$ vanish outside $\R_1$, the $|\nu^k\rangle$ vanish 
outside $\R_2$. Recall that $\R_1$ and $\R_2$ do not overlap. Finally, 
the normalization condition in~(\ref{inner-product}) follows from the 
unitarity of $\sigma_p$. 
In addition note that for $p \ne p'$ both belonging to $\P$,
\begin{equation}
\label{p-diagonal}
\langle\psi_{m,n,p}| \H_{N_1,N_2}^\C|\psi_{m',n',p'}\rangle=\langle 
\chi_n|\langle \phi_m|\sigma_p^{-1}\H_{N_1,N_2}^\C\sigma_{p'}| 
\phi_{m'}\rangle 
|\chi_{n'}\rangle=0,
\end{equation}
because the Hamiltonian does not interchange particles between the two regions
$\R_1$ and $\R_2$. 

We can  use the set of states $\{|\psi_{m,n,p}
\rangle\}$ to evaluate the trace in the partition function~(\ref{partfunc})
\begin{equation}
Z_{N_1,N_2}^{\C}={1 \over N!} \sum_{m,n} \sum_{p \epsilon \P}
\langle \chi_n|\langle \phi_m|\sigma_p^{-1}
~e^{-\H_{N_1,N_2}^{\C}/k_BT}~\sigma_p 
| \phi_m\rangle |\chi_n\rangle.
\end{equation}
The Hamiltonian, being symmetric, commutes with $\sigma_p$. The
sum over $p$  therefore just gives a factor of $P$, so that
\begin{equation}
Z_{N_1,N_2}^{\C}={1 \over N_1!N_2!} \sum_{m,n}
\langle \chi_n|\langle \phi_m|
~e^{-\H_{N_1,N_2}^{\C}/k_BT}~
| \phi_m\rangle |\chi_n\rangle,
\end{equation}
which is the same as the partition function $Z_{N_1,N_2}^{\C,\U}$ (defined in
equation~(\ref{unsym-partf})) evaluated 
in the Hilbert space spanned by $\{|\psi_{m,n}^\U\rangle\}$. The free energy 
$F_{N_1,N_2}^{\C}$ therefore is equal to $F_{N_1,N_2}^{\C,\U}$.

For fermions ($-$) and bosons ($+$) the appropriately symmetrized 
orthonormal states are
\begin{equation}
\label{fermi-bose}
|\psi_{m,n}^\pm\rangle= P^{-{1 \over 2}} \sum_{p\epsilon \P} (\pm)^{\pi(p)}
|\phi_m(1,...,N_1) \rangle~|\chi_n(N_1+1,...,N_1+N_2) \rangle,
\end{equation}  
where $\pi(p)$ is $0$ for an even and $1$ for an odd permutation $p$, 
and  $|\phi_m\rangle$ 
and $|\chi_n\rangle$ are assumed to have appropriate symmetry with respect to 
interchange of particles within $\R_1$ and within $\R_2$, respectively. We use 
the set of states $\{|\psi_{m,n}^\pm\rangle\}$ 
to evaluate the trace in the partition function~(\ref{partfunc}):
\begin{equation}
Z_{N_1,N_2}^{\C}=\sum_{m,n} {1 \over P} \sum_{p \epsilon \P} \sum_{p \epsilon 
\P}(\pm)^{\pi(p)+\pi(p')}\langle \chi_n|\langle \phi_m|\sigma_{p'}^{-1}
~e^{-\H/k_BT}~\sigma_p | \phi_m\rangle |\chi_n\rangle.
\end{equation}
Only terms with $p=p'$ survive because of~(\ref{p-diagonal}). Since 
$(\pm)^{2\pi(p)}=1$, the rest of the proof is similar 
to the Boltzmann case.

Step 3 in section~\ref{quantum}  follows from the observation that the
partition function $Z^\C$ for $N$ particles in $\R_1 \cup \R_2$ is a sum
of positive terms of the form
\begin{equation}
Z^{\C} = \sum_{N_1=0}^N Z_{N_1,N-N_1}^\C,
\end{equation}
where $Z_{N_1,N_2}^\C$ is the partition function for $N_1$ particles in $\R_1$
and $N_2$ particles in $\R_2$. Consequently $Z^{\C}$ is not smaller than
$Z_{N_1,N_2}^\C$ for any particular $N_1$ and $N_2=N-N_1$, and
\begin{equation}
\label{step3}
F^{\C}\le F_{N_1,N_2}^{\C}.
\end{equation}

Step 4 in section~\ref{quantum} exploits the fact that region $\R$
contains $\R_1 \cup \R_2$. Whenever
particles are allowed to move in a larger region, the partition function goes
up and the free energy goes down. The quantum version of this result is based
on the fact that an energy eigenstate $|\psi_m\rangle$ of $\H$ in the smaller
region, with energy $E_m$, belongs to the Hilbert space of the larger region.
Although it is not an eigenstate of $\H$ in the larger region, it is still
the case that
\begin{equation}
\label{property}
\langle \psi_m| \H | \psi_m \rangle= E_m.
\end{equation}
Now we make use of Peierl's theorem,
which states that
\begin{equation}
\label{theorem}
 {\rm Tr}~e^{-\H/k_B T}\ge  \sum_l e^{- \langle \phi_l| 
{\H/k_B T} |\phi_l \rangle}
\end{equation}
where $\{|\phi_l \rangle\}$ is any orthonormal set of states. Choose
$\{|\phi_l \rangle\}$ to be the set of energy eigenstates $\{|\psi_m\rangle\}$
of $\H$ in $\R_1 \cup \R_2$. It follows that the partition function $Z$ for 
region $\R$, is greater than $Z^\C$ for $\R_1 \cup \R_2$, and thus
\begin{equation}
F \le F^\C.
\end{equation}

\begin{figure}[tb]
\epsfxsize=6in \epsfbox{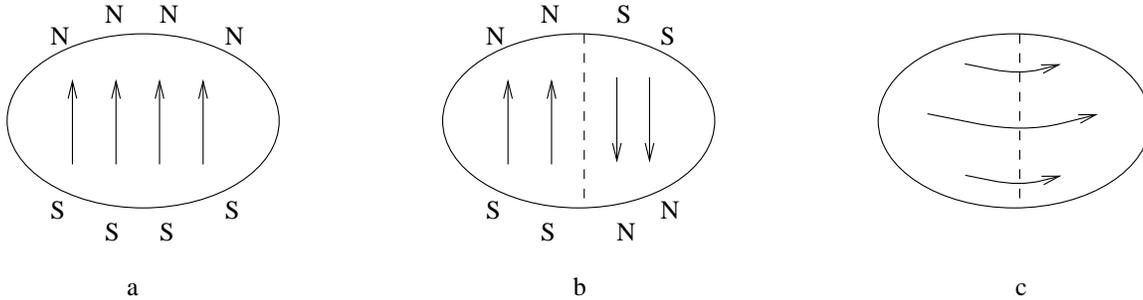}
\caption{(a) Surface poles due to uniform magnetization. 
(b) Magnetic domains separated by a domain wall. (c) Continuous
magnetization texture with vortex.}
\label{f1}
\end{figure}

\begin{figure}[tb]
\epsfxsize=6in \epsfbox{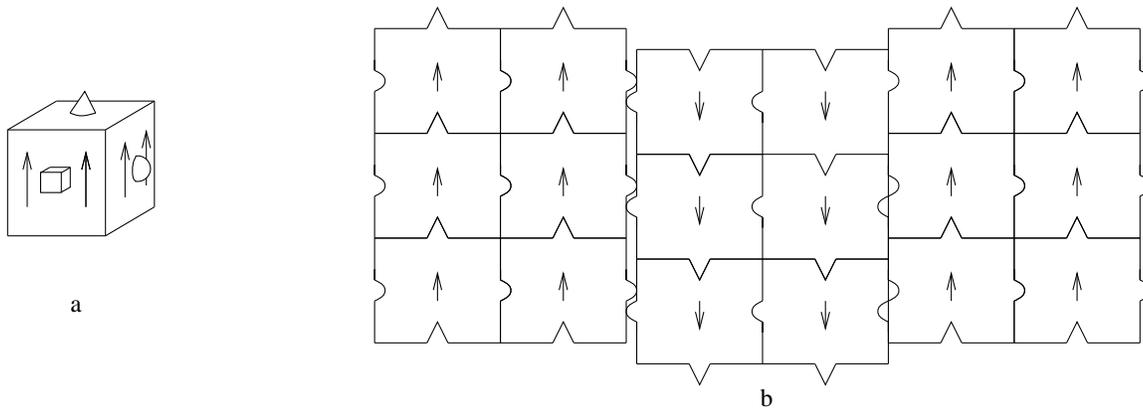}
\nopagebreak
\caption{(a) Permanently magnetized particle lacking a $\theta$ operator. The
arrows indicate the direction of magnetization. (b) Domain formation requires 
gaps between particles. A cross-section is shown.}
\label{f2}
\end{figure}

%\begin{table}[h]
%\label{summary}
%\caption{A list of models for which we have proved the existence of a shape
%independent thermodynamic limit}
%\begin{center}
%\begin{tabular}{||c|c|c|c||}\hline\hline
%Model & Type (Magnetic/Electric) & Proof section \\
%\hline \hline
%Dipolar hard spheres & Magnetic, Electric & 3, 5 \\
%\hline
%Polydisperse hard core particles & Magnetic, Electric & 4.1, 5 \\
%\hline
%Dipoles with central forces & Magnetic, Electric & 4.2, 5 \\
%\hline
%Quantum particles & Magnetic & 4.3 \\
%\hline
%Polarizable dipoles & Magnetic, Electric & 4.4, 5 \\
%\hline \hline
%\end{tabular}
%\end{center}
%\end{table}

\end{document}